\documentclass[twocolumn]{aastex61}
\usepackage{amsmath}
\usepackage{enumerate}
\usepackage{hyperref}
\usepackage{color}
\usepackage{subfigure}
\usepackage{color}

\received{}
\revised{}
\accepted{}
\submitjournal{}

\shorttitle{Spatially-resolved SFR-M$_{\ast}$ relation}
\shortauthors{Pan et al.}
\begin{document}

\title{SDSS IV M\MakeLowercase{a}NGA: Dependence of Global and Spatially-resolved  SFR-M$_{\ast}$ Relations on Galaxy Properties}

\email{hapan@asiaa.sinica.edu.tw}

\author{Hsi-An Pan}
\affil{Academia Sinica, Institute of Astronomy \& Astrophysics (ASIAA), P.O. Box 23-141, Taipei 10617, Taiwan}

\author{Lihwai Lin}
\affiliation{Academia Sinica, Institute of Astronomy \& Astrophysics (ASIAA), P.O. Box 23-141, Taipei 10617, Taiwan}

\author{Bau-Ching Hsieh}
\affiliation{Academia Sinica, Institute of Astronomy \& Astrophysics (ASIAA), P.O. Box 23-141, Taipei 10617, Taiwan}

\author{Sebasti\'an F. S\'anchez}
\affiliation{Instituto de Astronom\'ia, Universidad Nacional Auton\'oma de Mexico, A.P. 70-264, 04510, M\'exico, D.F., México}

\author{H\'ector Ibarra-Medel}
\affiliation{Instituto de Astronom\'ia, Universidad Nacional Auton\'oma de Mexico, A.P. 70-264, 04510, M\'exico, D.F., México}

\author{M\'ed\'eric Boquien}
\affiliation{Unidad de Astronomía, Universidad de Antofagasta, Avenida Angamos 601, Antofagasta 1270300, Chile}

\author{Ivan Lacerna}
\affiliation{Instituto Milenio de Astrof\'isica, Av. Vicu\~na Mackenna 4860, Macul, Santiago, Chile}
\affiliation{Instituto de Astrof\'isica, Pontificia Universidad Cat\'olica de Chile, Av. V.~Mackenna 4860, 782-0436 Macul, Santiago, Chile}

\author{Maria Argudo-Fern\'andez}
\affiliation{Unidad de Astronomía, Universidad de Antofagasta, Avenida Angamos 601, Antofagasta 1270300, Chile}

\author{Dmitry Bizyaev}
\affiliation{Apache Point Observatory and New Mexico State
	University, P.O. Box 59, Sunspot, NM, 88349-0059, USA}
\affiliation{Sternberg Astronomical Institute, Moscow State
	University, Moscow}

\author{Mariana Cano-D\'iaz}
\affiliation{Instituto de Astronom\'ia, Universidad Nacional Auton\'oma de Mexico, A.P. 70-264, 04510, M\'exico, D.F., México}

\author{Niv Drory}
\affiliation{McDonald Observatory, The University of Texas at Austin, 1 University Station, Austin, TX 78712, USA}

\author{Yang Gao}
\affiliation{Shanghai Astronomical Observatory, Chinese Academy of Sciences, 80 Nandan Road, Shanghai 200030, China}

\author{Karen Masters}
\affiliation{Institute of Cosmology and Gravitation, University of Portsmouth, Dennis Sciama Building, Burnaby Road, Portsmouth PO1 3FX, UK}

\author{Kaike Pan}
\affiliation{Apache Point Observatory and New Mexico State
	University, P.O. Box 59, Sunspot, NM, 88349-0059, USA}

\author{Martha Tabor}
\affiliation{School of Physics and Astronomy, University of Nottingham, University Park, Nottingham, NG7 2RD, UK}

\author{Patricia Tissera}
\affiliation{Departamento de Ciencias Fısicas, Universidad Andres Bello, Av. Republica 220, Santiago, Chile}

\author{Ting Xiao}
\affiliation{Shanghai Astronomical Observatory, Chinese Academy of Sciences, 80 Nandan Road, Shanghai 200030, China}
\affiliation{Department of Physics, Zhejiang University, Hangzhou 310027, China}



\begin{abstract} % 250 words %
%Galaxy  integrated H$\alpha$(global)-stellar mass(M$_{\ast}$(global)) relation (or star formation rate (SFR(global))-M$_{\ast}$(global) relation) 
Galaxy  integrated  H$\alpha$ star formation rate-stellar mass relation, or SFR(global)-M$_{\ast}$(global) relation,
is crucial for understanding  star formation history and evolution of galaxies.
However, many  studies have dealt with SFR using unresolved measurements, which makes it difficult to separate out the  contamination from other ionizing sources, such as active galactic nuclei (AGN) and evolved stars.
Using the integral field spectroscopic observations from SDSS-IV MaNGA, we spatially disentangle  the contribution from different H$\alpha$ powering sources for $\sim$1000  galaxies.
%We find that, when accounting for all ionizing sources in galaxies, the
We find that,  when including regions dominated by all ionizing sources in galaxies, the
spatially-resolved relation between H$\alpha$ surface density ($\Sigma_{\mathrm{H\alpha}}$(all)) and stellar mass surface density ($\Sigma_{\ast}$(all))  progressively turns over at high $\Sigma_{\ast}$(all) end for increasing  M$_{\ast}$(global)  and/or bulge dominance (bulge-to-total light ratio, B/T). This in turn leads to the  flattening of the integrated H$\alpha$(global)-M$_{\ast}$(global)  relation in the literature.
By contrast,   there is no noticeable flattening  in both integrated H$\alpha$(HII)-M$_{\ast}$(HII)  and  spatially-resolved $\Sigma_{\mathrm{H\alpha}}$(HII)-$\Sigma_{\ast}$(HII) relations  when only regions where star formation dominates the ionization are considered.
%when the star-forming  regions alone are considered.
In other words, the flattening can be attributed to the increasing regions powered by non-star-formation  sources, which generally have lower ionizing ability than star formation.
Analysis of the fractional contribution of  non-star-formation   sources to  total H$\alpha$ luminosity of a galaxy suggests a decreasing role of star formation as an ionizing source toward high-mass, high-B/T galaxies and bulge regions.
This result indicates that the  appearance of  the galaxy   integrated SFR-M$_{\ast}$ relation critically depends on their global properties (M$_{\ast}$(global) and B/T) and relative abundances of various ionizing sources within the  galaxies. 

\end{abstract}




\section{Introduction}
The relation between  galaxy star formation rate (SFR) and  stellar mass (M$_{\ast}$) provides key constraints on the star formation history and mass assembly of galaxies. Star-forming galaxies, populated by disk-dominated galaxies,  form a tight relationship on the SFR-M$_{\ast}$ plane, the  so-called ``star-forming main sequence'', which can be described by a power-law relation \citep{Noe07,Elb11,Cat15,Lee15}. On the other hand, the quiescent population, primarily composed of bulge-dominated galaxies \citep[e.g.,][]{Wuy11}, has  a much lower specific star formation rate (sSFR $\equiv$ SFR/M$_{\ast}$) with respect to the main sequence. Several  studies have suggested that the main sequence relation flattens at high mass end, possibly due to the growth of the bulge that lower the global sSFR of a galaxy \citep{Noe07,Abr14,Whi15,Cat17}.

Although the  SFR-M$_{\ast}$ relation has been reported with  a variety of different data sets, their  appearance can vary  significantly from one  to another. 
A key uncertainty occurs in the SFR measurement.
H$\alpha$ is  frequently utilized as SFR indicators.
However, it has long been known  that young star  is not the only powering source of H$\alpha$.
Other  sources  such as active galactic nucleus (AGN) and old stellar population  also  contribute to H$\alpha$ \citep{Yan12,Sin13,Bel16} and therefore  affect the  SFR-M$_{\ast}$ relation. 

Since star formation is an intrinsically local process, 
further insight into the nature of SFR-M$_{\ast}$ relation  would require spatially-resolved  data to understand what processes drive the connection between SFR and M$_{\ast}$.
Using spatially-resolved spectroscopy, recent studies have shown  that there also exist a tight correlation between   SFR and stellar mass surface density  for nearby and distant star-forming galaxies \citep{Nel12,Wuy13,Can16,Abd17,Hsi17,Ell17}. 
Recently, \cite{Hsi17}  extends the study to the quiescent population using the MaNGA survey \citep[Mapping Nearby Galaxies at APO;][]{Bun15} and find that the emission line fluxes, even classified as LI(N)ER\footnote{Low-ionization (nuclear) emission line regions. There is growing evidence   that LI(N)ER  is not	exclusively  powered by  the central AGN, but  also ionizing  sources in  galactic disk  \citep[e.g.,][]{Bel16}. }, are also correlated with the underlying stellar mass surface density. Therefore, it is essential to separate out the contribution of non-star-forming regions when measuring the  SFR based on emission line methods. This also has an important application to galaxy formation models as the SFR-M$_{\ast}$ relation is often used to validate the subgrid physics modeling \citep[][]{Tis16,Lag16}.

The main aim of this paper is to show that  the galaxy SFR-M$_{\ast}$ relation is sensitive to whether or not the ``contamination'' is removed.
Since H$\alpha$ does not only trace  star formation, throughout the paper we refer  to SFR-M$_{\ast}$ relation as \emph{H$\alpha$-M$_{\ast}$ relation}. 
SFR axis and  sSFR lines are  provided  for  readers to compare with other studies.

We present our study as follows:  in Section \ref{sec_data}, we describe our sample and data, and present the traditional global H$\alpha$-M$_{\ast}$ relation making use of the total H$\alpha$ luminosity and M$_{\ast}$ of galaxies.  Section \ref{sec_results} compares the spatially-resolved H$\alpha$-M$_{\ast}$ surface density relation before and after the non-star-formation  sources being removed.
Section \ref{sec_origin} quantifies the contribution of the non-star-formation  source as a function of galaxy properties and compares the integrated H$\alpha$-M$_{\ast}$ relations with and without  the non-star-formation contributions being removed.  Main  results are summarized in Section \ref{sec_summary}.

\section{Data and the Traditional Global H$\alpha$-M$_{\ast}$ Relation}
\label{sec_data}
\subsection{MaNGA Survey}

The advent of MaNGA survey \citep{Bun15,Law15,Yan16a}, which spatially resolves stellar and gas properties, offers an excellent opportunity to examine the H$\alpha$-M$_{\ast}$ relation.
MaNGA   is part of the fourth generation of the Sloan Digital Sky Survey \citep[SDSS-IV;][]{Gun06,Bla17} and aims to obtain spatially-resolved spectroscopy of 10,000 galaxies  with  median redshift $\sim$ 0.03 by 2020.
Further details on the MaNGA sample selection can be found in \cite{Wak17}.
MaNGA has a wavelength coverage of 3600 -- 10300 \AA, with a spectral resolution varying from $R$ $\sim$ 1400 at 4000 \AA\,  to $R$ $\sim$ 2600 around 9000 \AA\,\citep{Sme13,Yan16b}.
MaNGA uses 5 different types of IFU, ranging in diameter from 19 (12.5$\arcsec$)  to 127 fibers  (32.5$\arcsec$). 
The IFUs are installed in six SDSS cartridges.
Each MaNGA cartridge has 17 science IFUs\footnote{The MaNGA science IFU complement is 2 $\times$ 19-fiber IFU,  4 $\times$ 37-fiber IFU, 4 $\times$ 61-fiber IFU,  2 $\times$ 91-fiber IFU,   and 5 $\times$ 127-fiber IFU per cartridge.} and 12 seven-fiber IFUs for calibration.
The IFU sizes and the number density of galaxies on the sky were designed jointly to allow more efficient use of IFUs (e.g.,  minimize the number of IFUs that are unused due to tile with too few galaxies), and to allow  to observe galaxies in the redshift range to at least 1.5 effective radii \citep{Dro15,Wak17}.

This study draws data from the fourth MaNGA Product Launches (MPL-4), corresponding to  SDSS DR13 \citep{sdssdr13}.
The observational data was reduced using the MaNGA data-reduction-pipeline \citep[DRP;][]{Law16}.

\subsection{Local and Global SFR and M$_{\ast}$ Measurements}
\label{sec_measurements}
The reduced spaxel-wise  data cubes were analyzed using the Pipe3D pipeline  to extract the  physical parameters from each of the spaxels in each galaxy.
Pipe3D fits the continuum with stellar population models and measures the nebular emission lines. Details of the  procedures and uncertainties of the process are described in \citet{San16a,San16b} and \citet{San17}. 

We briefly summarize the fitting of stellar continuum and the derivation of emission line flux here.
The stellar continuum was first modeled using a simple-stellar-population (SSP) library  with 156 SSPs, comprising 39 ages and 4 metallicities \citep{Cid13,San16b}. 
Before the fitting,  spatial binning is  performed to reach a signal-to-noise ratio (S/N) of 50 across the field of view. 
Then  the stellar population fitting was applied to the coadded spectra within each spatial bin.
Finally, the stellar population model for spaxels with continuum S/N $>$ 3 is derived by re-scaling the best fitted model within each spatial bin to the continuum flux intensity  in  the  corresponding spaxel. 
The stellar mass  is obtained using the stellar populations derived for each spaxel, then normalized to the physical area of one spaxel to get the surface density ($\Sigma_{\ast}$) in unit of M$_{\sun}$ kpc$^{-2}$. The stellar mass per spaxel is also coadded to derive the integrated stellar mass of the galaxies (M$_{\ast}$(global)).

The stellar-population model  are  subtracted from the data cube to create an ionized gas emission line cube (with noise). 
The emission line fluxes were measured spaxel by spaxel.
The  SFR  was derived using the  H$\alpha$ emission line.  
It is again possible to compute the total H$\alpha$ luminosity (H$\alpha$(global)) and SFR (SFR(global)) by integrating the   spatially-resolved quantities over  spaxels.
The integrated H$\alpha$ luminosity  were derived using the H$\alpha$
fluxes for all the spaxels with  S/N $>$ 3. 
%We have done several tests by  doing a cut in higher S/Ns.
%The integrated H$\alpha$ luminosity does not change, since it is dominated by the signal in those spaxels with clearly high S/N.\\

To study the effect of non-star formation powered H$\alpha$  on the  H$\alpha$-M$_{\ast}$ relation, we  use a set of  emission line ratios to spatially distinguish the ionization mechanisms of H$\alpha$ in galaxies (see \S\ref{sec_resv_hii}). 
To ensure reliable emission line ratios, we also limit the spatially-resolved analysis to spaxels\footnote{Hereafter, the term spaxel refers to only spaxels with S/N $>$ 3 in the emission line fluxes and continuum used.} with S/N(H$\alpha$), S/N(H$\beta$), S/N([\ion{O}{3}]), and S/N([\ion{N}{2}]) $>$ 3. %This, again, has minimal impact on our results as the results are dominated by the high-S/N spaxels.
The fluxes are converted to luminosities and corrected for extinction.
The method described in the Appendix of \cite{Vog13} is used to compute the reddening using the Balmer decrement at each spaxel of the IFU cube.  
The extinction-corrected H$\alpha$ luminosity is converted into SFR surface density  ($\Sigma_\mathrm{SFR}$ in M$_{\sun}$ yr$^{-1}$ kpc$^{-2}$) using the empirical calibration from \cite{Ken98} that adopts the Salpeter IMF.

Inclination correction  is  applied to the H$\alpha$ luminosity and stellar mass of all spaxels of a galaxy equally.
Galaxy inclination measured by the disk ellipticity in \cite{Sim11}  is adopted (see the next section). 
Such  correction is based on the assumption of thin disks, whereas for round bulges or more  spheroidal galaxies it may systematically overestimate the effect of projection  and thus underestimate the H$\alpha$ luminosity and stellar mass.
However, we also note that the correction does not affect the sSFR related quantities since it is applied to both  H$\alpha$ luminosity and stellar mass.

MaNGA galaxies  are selected to have spectroscopic coverage to 1.5 -- 2.5 effective radii ($R_{e}$). The exact range varies from galaxy to galaxy.
The mean offset between the Pipe3D M$_{\ast}$(global) and  the aperture-corrected NSA\footnote{The NASA-Sloan Atlas: http://nsatlas.org} stellar mass is 0.07 dex, corresponding to the difference between the adopted cosmologies and the differences in IMFs. 
We then assume that the aperture effect has minimal impact on the stellar mass.
To estimate  how much of the SFR in a galaxy may be unaccounted for due to the   finite fiber aperture, we assume that H$\alpha$ following a simple exponential profile  out to infinite radius. 
The  disk exponential profile in $r$-band of each galaxy derived by \citet{Sim11} is adopted.
We estimate the total SFR out to infinite radii and SFR inside the MaNGA IFUs for all galaxies, and find that most of the star formation (on average, 84 $\pm$ 15\%) occurs within the MaNGA IFUs. 
In light of this,  we assume that the aperture effect does not significantly affect the SFR measurement as well.

MaNGA targets  galaxies in the redshift range 0.01 $<$ $z$ $<$ 0.15  \citep{Wak17}. 
Since the  spatial resolution is different across the sample, we	repeat the analysis in this work by using the sub-samples selected by  distance. The main results are not severely affected by the varying physical size of spaxels.

\subsection{The Bulge-Disk Decomposition}
\label{sec_decomposition}
Galaxy structural parameters are taken from the bulge-disk decomposition  catalog from \cite{Sim11}.
\cite{Sim11} perform the two-dimensional bulge and disk decompositions using the GIM2D software package \citep{Sim02} on the \emph{g}-band and \emph{r}-band  images of SDSS DR7 galaxies. 
In the model,  the bulge S\'ersic index ($n$) is treated as a  free parameter and the disk component has $n$ $=$ 1.
 Structural parameters measured in \emph{r}-band are used in this work.
We use the bulge-to-total light ratio (B/T)  as a proxy for bugle dominance.
The bulge and disk regions are separated by the radius at which  50\% of the light is contributed by the bulge and disk component respectively. Specifically, for each galaxy, we look for the intersection of the one-dimensional fractional \emph{r}-band  S\'ersic profile of  bulge and the exponential  profile of disk.
It must be noted that the radius does not indicate the physical size of the bulge, but the boundary of the bulge-\emph{dominated} and the disk-\emph{dominated} regions.

The sample has been selected to only include galaxies that have measurements from both Pipe3D and \cite{Sim11}. With this requirement, 
1037 out of $\sim$1400 galaxies in  MPL-4 are left.

\subsection{Traditional Global H$\alpha$-M$_{\ast}$ Relation}

\begin{figure}%

		\includegraphics[width=0.45\textwidth]{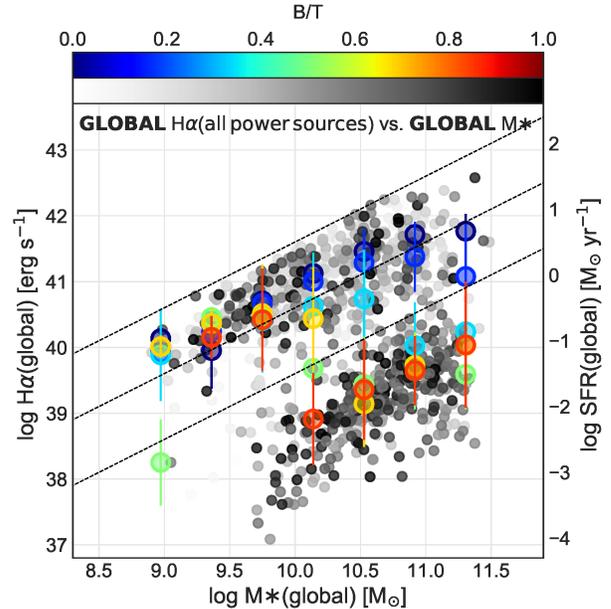}
	\caption{Integrated H$\alpha$(global)-M$_{\ast}$(global) relation for  individual galaxies (small circles) derived from the Pipe3D analysis, color-coded  by their B/T from white to black. Colored circles are the median values of log(H$\alpha$(global)/erg s$^{-1}$) of the whole sample in different M$_{\ast}$ and B/T bins, color-coded  by B/T (following the scheme of Fig. 2 in \cite{Whi15}). The error bars are given by the standard deviation in each bin. The dashed lines represent  log(sSFR/yr$^{-1}$) of -9.5, -10.5, and -11.5  (from top to bottom).  }%
	\label{fig_global_sm_sfr_integrated}%
\end{figure}

Figure \ref{fig_global_sm_sfr_integrated} shows the H$\alpha$(global)-M$_{\ast}$(global) relation using total H$\alpha$ luminosity and M$_{\ast}$ of galaxies.
The small circles  present the individual galaxies   color-coded by B/T from white to black. 
The dashed lines denote  log(sSFR/yr$^{-1}$) of -9.5, -10.5, and -11.5  (from top to bottom).
As reported in the literature, galaxies   populate two distinct sequences, with a clear separation between  star-forming and quiescent galaxies. 

To characterize the dependence of the H$\alpha$(global)-M$_{\ast}$(global) relation on B/T and M$_{\ast}$(global), we  binned the galaxies by these two quantities.
Big circles are the median values of H$\alpha$(global) of the whole sample in different M$_{\ast}$(global) bins, colored according to  B/T (following the  scheme of Fig. 2 in \cite{Whi15}). The discontinuity in the median values of H$\alpha$(global) at log(M$_{\ast}$(global)/M$_{\odot}$) $\sim$ 10 is caused by decreasing number of quiescent targets in the low mass end.

As  been noticed by many authors \citep[e.g.,][]{Wuy11}, the    sequence with lower H$\alpha$(global)-to-M$_{\ast}$(global) ratio, i.e., lower sSFR(global),  is occupied prevalently by bulge-dominated galaxies  (B/T $\geq$ 0.2), whereas the star-forming sequence is composed of   all populations  (but note that  disk-dominated galaxies with B/T $<$ 0.2 appear to be almost exclusively star-forming galaxies).
A flattening of the   lower-B/T galaxies ($\sim$ 0.2) at  log(M$_{\ast}$(global)/M$_{\odot}$) $>$ 11 is observed.
This  has been explained as the increasing fraction of the mass being given by bulges that have begun to quench, indicating a transition from disk to bulge-dominated properties.

\section{Spatially-Resolved H$\alpha$-M$_{\ast}$ Relation}
\label{sec_results}

\subsection{Spatially-Resolved H$\alpha$-M$_{\ast}$ Relation Using All Spaxels in Galaxies}
\begin{figure*}%
	\subfigure[]{%
		\includegraphics[width=0.48\textwidth]{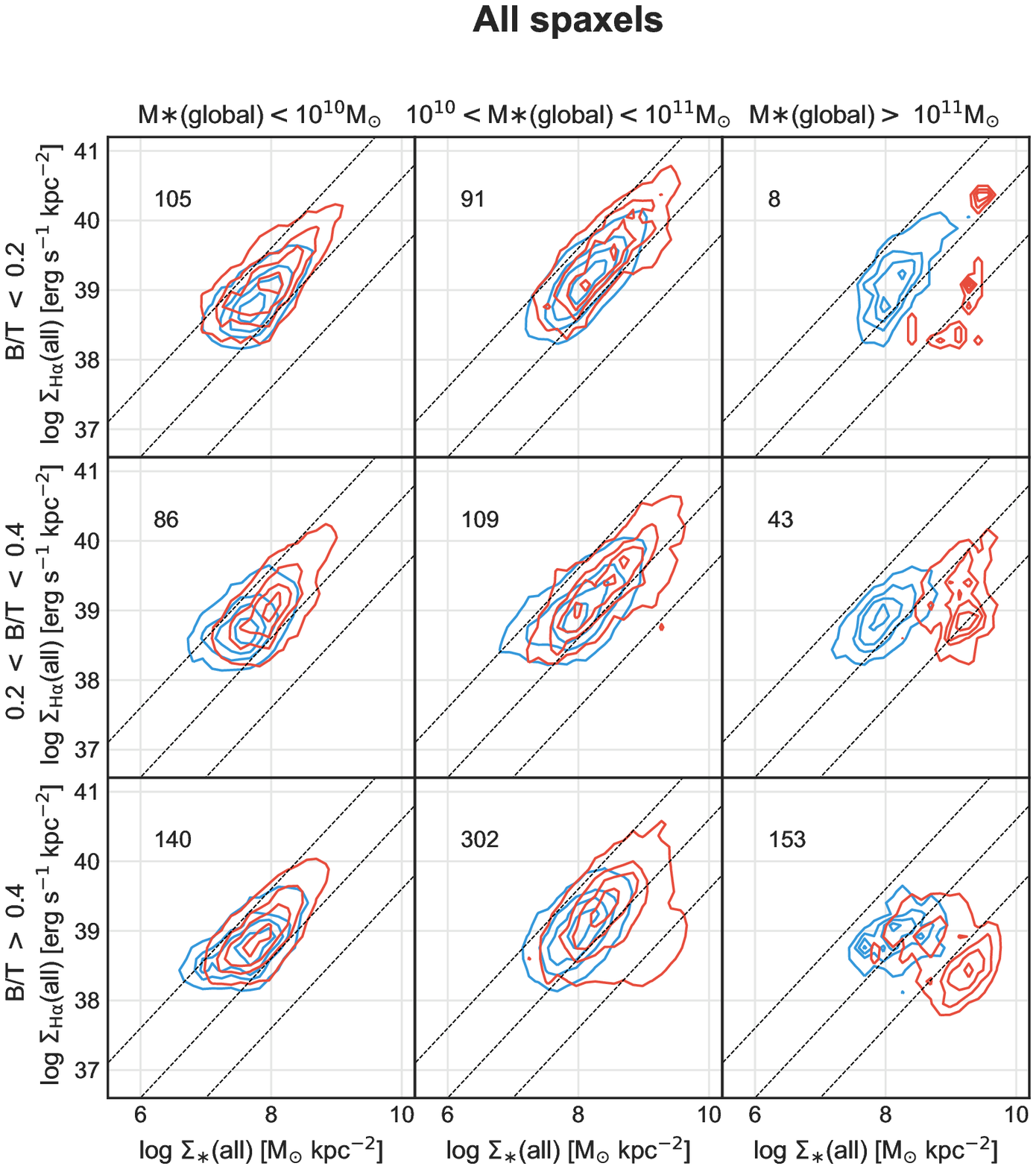}
		\label{fig_sfr_sm_BPTall}}
	\subfigure[]{%
		\includegraphics[width=0.48\textwidth]{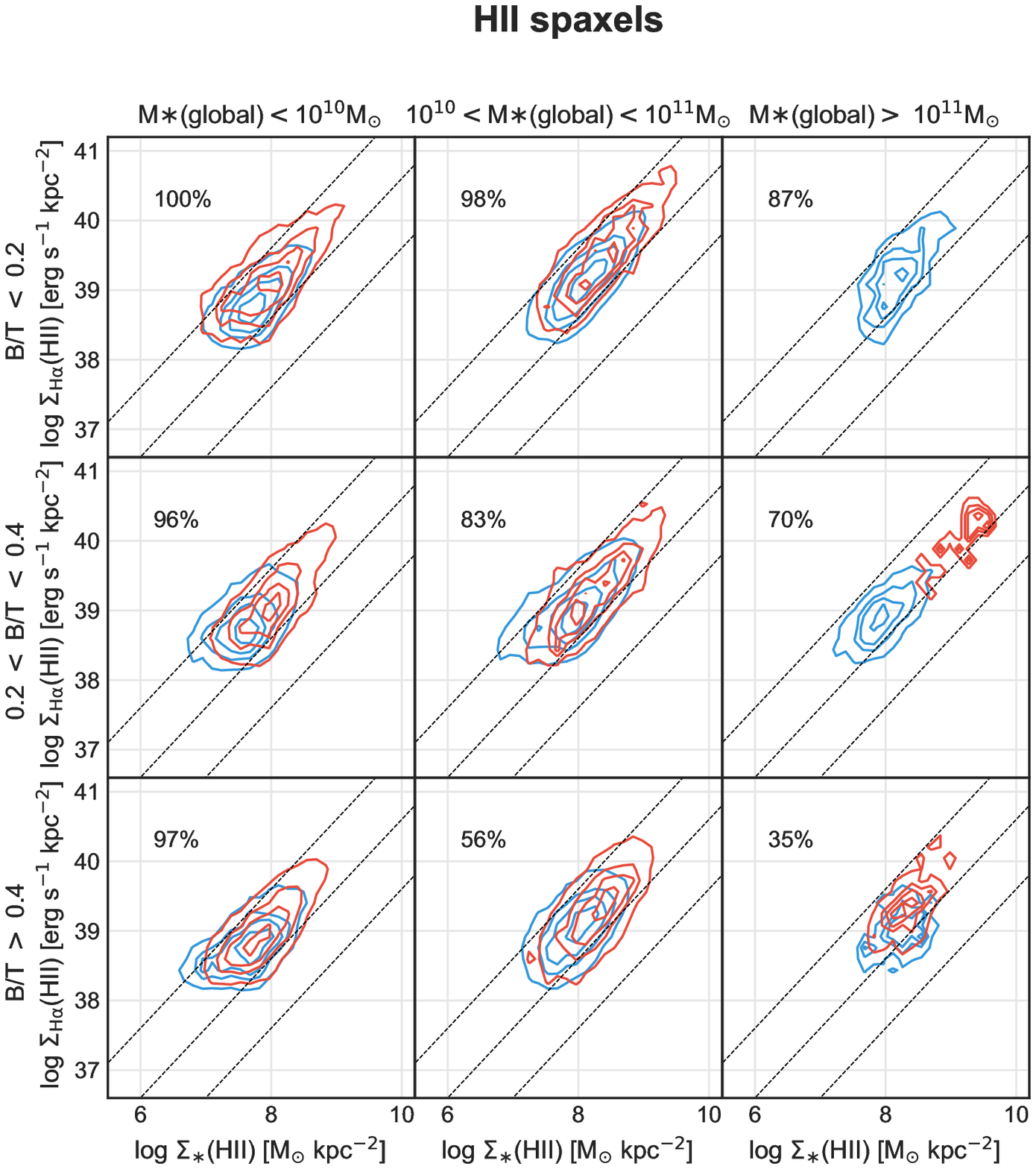}
		\label{fig_sfr_sm_BPThii}
	}
	\caption{(a) Inclination-corrected spatially-resolved $\Sigma_{\mathrm{H\alpha}}$(all)-$\Sigma_{\ast}$(all) relation. All spaxels  (S/N $>$ 3)  powered by all ionizing mechanisms  in the galaxies are used to make the plot. Blue and red contours denote the spaxels from  disk and bulge, respectively. B/T and M$\ast$(global) increase from the top to  bottom  and the left to  right. In other words, from  upper-left to  bottom-right sub-panel, galaxies go from disk-dominated to bulge-dominated. Number of galaxy in each  bins is indicated at the upper-left corner of the sub-panels. The contours represent 15, 40, 60, and 85\% of the peak counts, per 0.15 dex-wide cell.   The dashed lines denote  log(sSFR/yr$^{-1}$) of -9.5, -10.5, and -11.5 (from top to bottom). (b) Inclination-corrected  $\Sigma_{\mathrm{H\alpha}}$(HII)-$\Sigma_{\ast}$(HII) relation. Only HII spaxels (i.e. star-forming regions) are used. The percentage value in the upper-left corner of each sub-panel indicates the number fraction of galaxies with HII spaxels relative to the number of galaxies in each bin in panel (a). }%
	\label{fig_resolved_sm_sfr}%
\end{figure*}

The spatially-resolved  $\Sigma_{\mathrm{H\alpha}}$ and  $\Sigma_{\ast}$ maps of MaNGA allow us to probe the driver of the   H$\alpha$(global)-M$_{\ast}$(global) relation in more detail.
Figure \ref{fig_sfr_sm_BPTall} presents the  inclination-corrected spatially-resolved $ \Sigma_{\mathrm{H\alpha}}$(all)-$\Sigma_{\ast}$(all) relation using all spaxels  of galaxies. 
The galaxies are binned by their   M$_{\ast}$(global) and B/T: from the upper-left to the bottom-right sub-panel, galaxies go from disk-dominated to bulge-dominated.
Number of galaxy in each bins is indicated in the upper-left corner.
Bulge and disk regions are shown by red and blue contours, respectively.  
Number of spaxel in each sub-panel ranges from $\sim$ 900 to 39000 for the bulge  and $\sim$ 6500 to 70000 for the disk.

For galaxies with   log(M$_{\ast}$(global)/M$_{\odot}$) $<$ 10,     bulge and disk lie along a similar $ \Sigma_{\mathrm{H\alpha}}$(all)-$\Sigma_{\ast}$(all) relation.
As the stellar mass increases to 10 $<$ log(M$_{\ast}$(global)/M$_{\odot}$) $<$ 11,  the high-mass ends of  bulge sequence start to move downward (i.e., decrease in  the $\Sigma_{\mathrm{H\alpha}}$(all)-to-$\Sigma_{\ast}$(all) ratio).
The decrease is more pronounced in the high-B/T galaxies than in the low-B/T galaxies.
In the most massive galaxies (log(M$_{\ast}$(global)/M$_{\odot}$) $>$ 11), the entire bulge sequence drops below the relations of the lower-mass objects.  
Meanwhile, the disk sequence also shows a slight drop in the $ \Sigma_{\mathrm{H\alpha}}$(all)-to-$\Sigma_{\ast}$(all) ratio at the  log(M$_{\ast}$(global)/M$_{\odot}$) and B/T values above 10 and 0.4, respectively (lower-right sub-panel).
The combination of these   leads to the  high-M$_{\ast}$(global) and/or high-B/T  systems being pulled  off the main sequence on the integrated H$\alpha$(global)-M$_{\ast}$(global) plane. 
Such turnover also indicates the  inside-out quenching of galaxies. The quenching process is beyond the scope of the present paper, but is of major importance in the context of galaxy evolution \citep[e.g.,][]{Li15,Gon16,Bel17,Lin17,Ell17,Spi17}.

\subsection{Spatially-Resolved H$\alpha$-M$_{\ast}$ Relation Using Star-Formation Spaxels in Galaxies}
\label{sec_resv_hii}

We now turn our attention to  the powering source of H$\alpha$. It is known that
  massive stars are not the only  source capable of providing  ionizing photons. %massive stars are not the only  source capable of emitting large quantities at H$\alpha$. 
To disentangle  different powering sources, we use the   emission line ratio diagnostics, the  BPT diagram \citep{Bal81} and H$\alpha$ equivalent width (EW) to spatially identify the regions ionized by different physical processes in each galaxy.
The emission line regions are classified into   star-forming HII regions,  LI(N)ER, Seyfert, and composite regions (mix of multiple  sources) based on their locations on the [\ion{O}{3}] 5007/H$\beta$ versus [\ion{N}{2}] 6584/H$\alpha$ plane \citep{Kew01,Kau03,Cid10}. In addition, the criterion of EW $>$ 6\AA\,  is also applied when selecting star-forming regions \citep{San14,San17}.

Armed with the spatially-resolved ionization sources of each galaxy, we use the identified  HII spaxels to construct the  $ \Sigma_{\mathrm{H\alpha}}$(HII)-$\Sigma_{\ast}$(HII) relation  driven by star formation alone.
The result is presented in Figure \ref{fig_sfr_sm_BPThii}.
For the star-forming regions, $\Sigma_{\mathrm{H\alpha}}$(HII) and $\Sigma_{\ast}$(HII) is much more tightly correlated than  that including all ionized regions of the galaxies.
Moreover, the bulge sequence shifts upwards to be close to   that of disk.
In other words, at least for the star-forming regions, the $ \Sigma_{\mathrm{H\alpha}}$(HII)-to-$\Sigma_{\ast}$(HII) ratio of bulge and disk, which is proportional to the local sSFR, do not differ significantly from each other.
In light of this, the turnover seen in the $ \Sigma_{\mathrm{H\alpha}}$(all)-$\Sigma_{\ast}$(all) relation  can be attributed to  non-HII regions; moreover,  the  non-HII ionizing sources tend to generate lower  H$\alpha$ luminosity than  that of star-forming regions and the difference can vary by up to an order of magnitude.  
Therefore, the total H$\alpha$ luminosity of a galaxy strongly depends on the relative proportion between HII and non-HII regions.

Another notable feature in Figure  \ref{fig_sfr_sm_BPThii} is the lack of HII spaxles  towards  higher masses and higher B/T.
The  number fraction of galaxies \emph{with} HII spaxels relative to the total number of galaxies in each bin is given in the upper-left corner of each sub-panel. 
The fraction is generally inversely correlated with M$_{\ast}$(global) and B/T, suggesting a decreasing role of  star formation as an ionizing source towards high-mass and high-B/T galaxies.

\section{Revisit the Integrated Relations}
\label{sec_origin}
The previous section indicates that the flattening of the spatially-resolved  bulge sequence can be attributed to the non-HII sources, which generally have lower ionizing ability compared to young stars, and such contribution become more significant with increasing M$_{\ast}$(global) and B/T. It is therefore worth quantifying the  contribution of non-HII powering sources in different galaxy populations and sub-galactic structures, and  revisiting the integrated H$\alpha$-M$_{\ast}$ relation of galaxies.

 \subsection{Quantitative Contribution of Non-HII Powering Sources}
 \label{sec_barplots}
\begin{figure*}%
	\subfigure[]{%
		\includegraphics[width=0.95\textwidth]{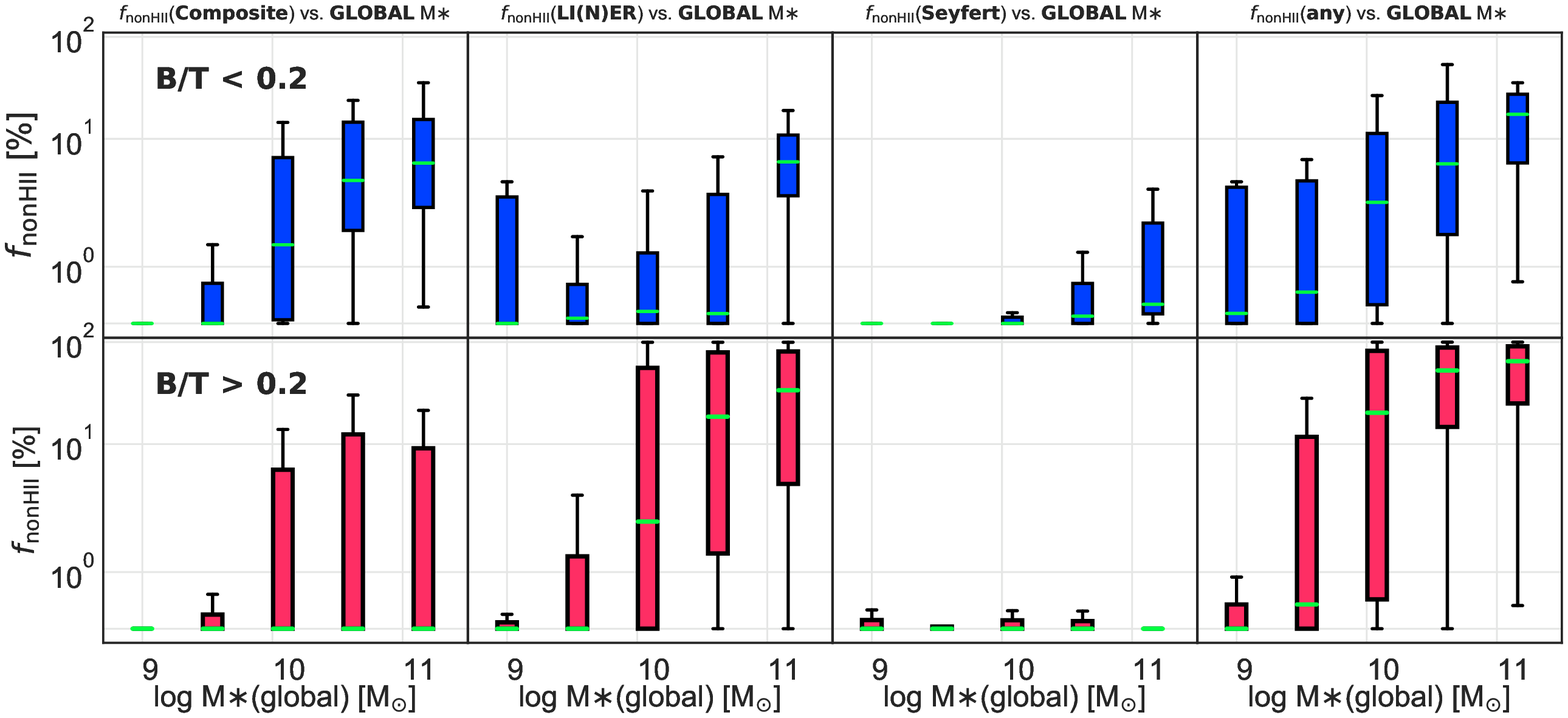}
		\label{fig_Frac_all}}
	\subfigure[]{%
		\includegraphics[width=0.95\textwidth]{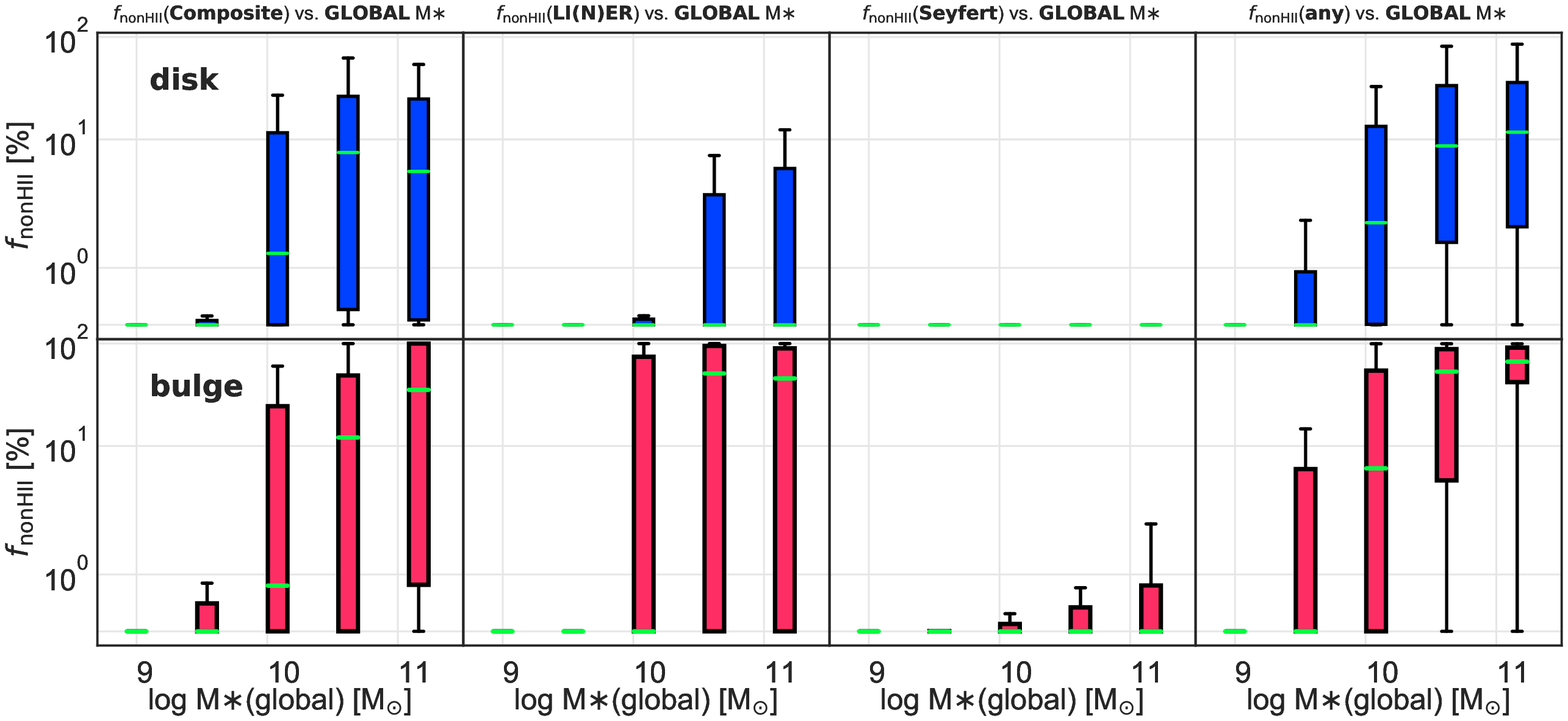}
		\label{fig_Frac_all3}}
	\caption{Box plots showing the $f_{\mathrm{nonHII}}$ distribution  for each of the M$_{\ast}$(global) bin.  In each plot the median is indicated by the green bar in the middle. The ends of the box are the upper and lower quartiles (the interquartile range, IQR),  50\% of the sample is located in the box. The two whiskers (vertical lines) outside the box extend to 1.5 $\times$ IQR,  99\% of the sample is inside the caps of the whiskers.  In some stellar mass bins,  the median and the  whiskers are subsumed in a single location due the large number of galaxies with  small $f_{\mathrm{nonHII}}$.  Panel (a)  shows  $f_{\mathrm{nonHII}}$ in each B/T for each M$_{\ast}$(global). The upper row  shows the results for galaxies with B/T $<$ 0.2 (disk-dominated galaxies), the lower row  shows galaxies with B/T $>$ 0.2 (bulge-dominated galaxies). From left to right, the sub-panels present the  $f_{\mathrm{nonHII}}$ of composite, LI(N)ER, and Seyfert, respectively. 
		Panel (b) shows the dependency of $f_{\mathrm{nonHII}}$ on galactic sub-regions, with disks in the upper row and bulges in the lower row).   
	}
	\label{fig_Frac_all_all}
\end{figure*}

 The  box plots in Figure \ref{fig_Frac_all_all}  describe the distribution of  the fraction of non-HII  contribution in the total H$\alpha$ luminosity of a galaxy  ($f_{\mathrm{nonHII}}$) in different  M$_{\ast}$(global) bins. 
 Three columns from left respectively present the fraction of H$\alpha$ contributed by composite, LI(N)ER, and  Seyfert, respectively, e.g., in the left-most column,  $f_{\mathrm{nonHII}}$ $=$ H$\alpha$(composite)/H$\alpha$(global).  
 The green line  drawn across the box is the sample median. 
 The ends of the box are the upper and lower quartiles (the interquartile range, IQR), i.e., 50\% of the sample is located in the box.
 The two whiskers (vertical lines) outside the box extend to 1.5 $\times$ IQR, i.e.,   99\% of the sample is inside the caps of the whiskers.
 In the following paragraphs, we will discuss    $f_{\mathrm{nonHII}}$  as a function of M$_{\ast}$(global), B/T, and galactic sub-structures.

 Figure \ref{fig_Frac_all} presents  the dependence of $f_{\mathrm{nonHII}}$ on the bulge dominance, B/T.  
 The upper and lower rows show the results for B/T $<$ 0.2 and $>$ 0.2, respectively. 
 Several features are readily apparent.
 Most notably,    the (non-zero) median $f_{\mathrm{nonHII}}$ increases in general with   increasing M$_{\ast}$(global) in both populations,  suggesting that the non-HII sources become more important with increasing M$_{\ast}$(global) \citep[see also][]{Cat17}.
For some M$_{\ast}$(global) bins,  the median and the  whiskers are subsumed in a single location due the large number of galaxies with  small $f_{\mathrm{nonHII}}$.

 Moreover, in the bulge-dominated galaxies, the non-HII contribution is exclusively dominated by LI(N)ER, whereas the three mechanisms  all make a certain  contribution, but typically lower than  LI(N)ER in the high-B/T galaxies,  in the disk-dominated galaxies.
  This originates  from the fact that the old stellar population, such as post-AGBs stars, have become the  main source of ionizing photons after star formation has ceased \citep{Yan12,Sin13,Bel16,Hsi17}.
As a whole, the right-most column shows that $f_{\mathrm{nonHII}}$ increases from  less than a percent   for log(M$_{\ast}$/M$_{\odot}$) $<$ 10 to a few to several tens percent  at log(M$_{\ast}$/M$_{\odot}$) $>$ 10.
Besides, high-B/T galaxies generally display a higher, or just comparable,  median $f_{\mathrm{nonHII}}$ to the low-B/T galaxies over the all range of masses.

 Figure \ref{fig_Frac_all3} explores the dependence of $f_{\mathrm{nonHII}}$ on  sub-galactic regions.
 The upper and lower rows show the disk and bulge regions, respectively. 
Bulges generally exhibit higher fraction of non-HII contribution than the disks, and the non-zero median $f_{\mathrm{nonHII}}$ increases with increasing M$_{\ast}$(global).
When accounting for all non-HII mechanisms (the right-most column), median $f_{\mathrm{nonHII}}$ is no higher than 20\% (mostly below 10\%) for  disks across all stellar mass bins, and increases to several tens percent in  bulges when  log(M$_{\ast}$/M$_{\odot}$) exceeding 10.
 The result is consistent with the study based on a 2D spectral decomposition of the bulge and disk component \citep{Cat17}.
 The high $f_{\mathrm{nonHII}}$ of bulge  is presumably due to the fact that the non-HII  sources (e.g., AGNs, evolved stars, and shocks) are naturally found most often in this old  central component.
 The above-mentioned characteristics   echo the  turnover feature of   bulge  $ \Sigma_{\mathrm{H\alpha}}$(all)-$\Sigma_{\ast}$(all) relation towards higher masses and higher B/T in Figure \ref{fig_sfr_sm_BPTall}.

\subsection{Revisiting the Integrated  H$\alpha$-M$_{\ast}$ Relation}
\label{sec_integrated}
\begin{figure*}%
	\subfigure[]{%
		\includegraphics[width=0.33\textwidth]{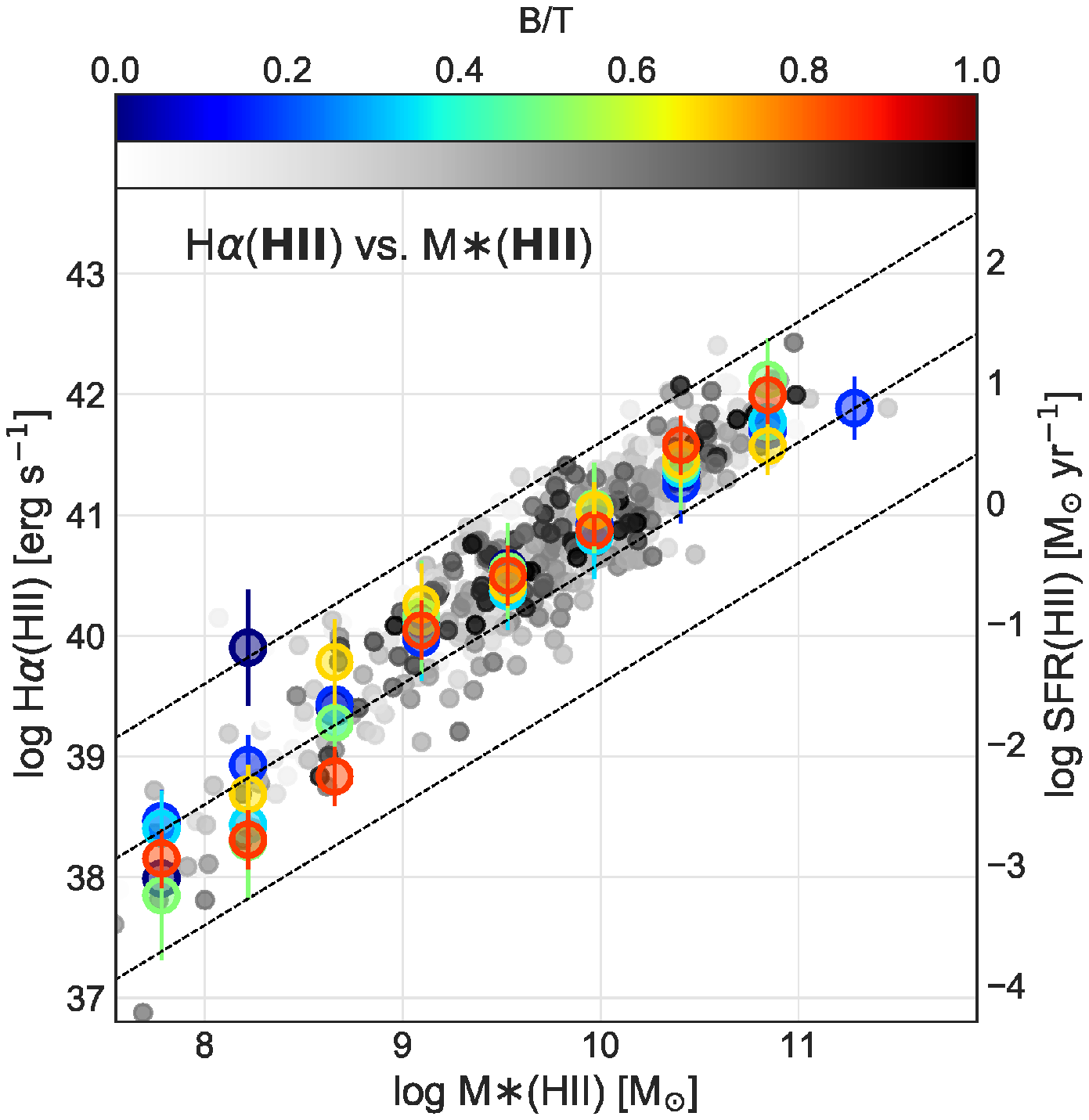}
		\label{fig_global_sm_sfr_hiih2}}
	\subfigure[]{%
		\includegraphics[width=0.33\textwidth]{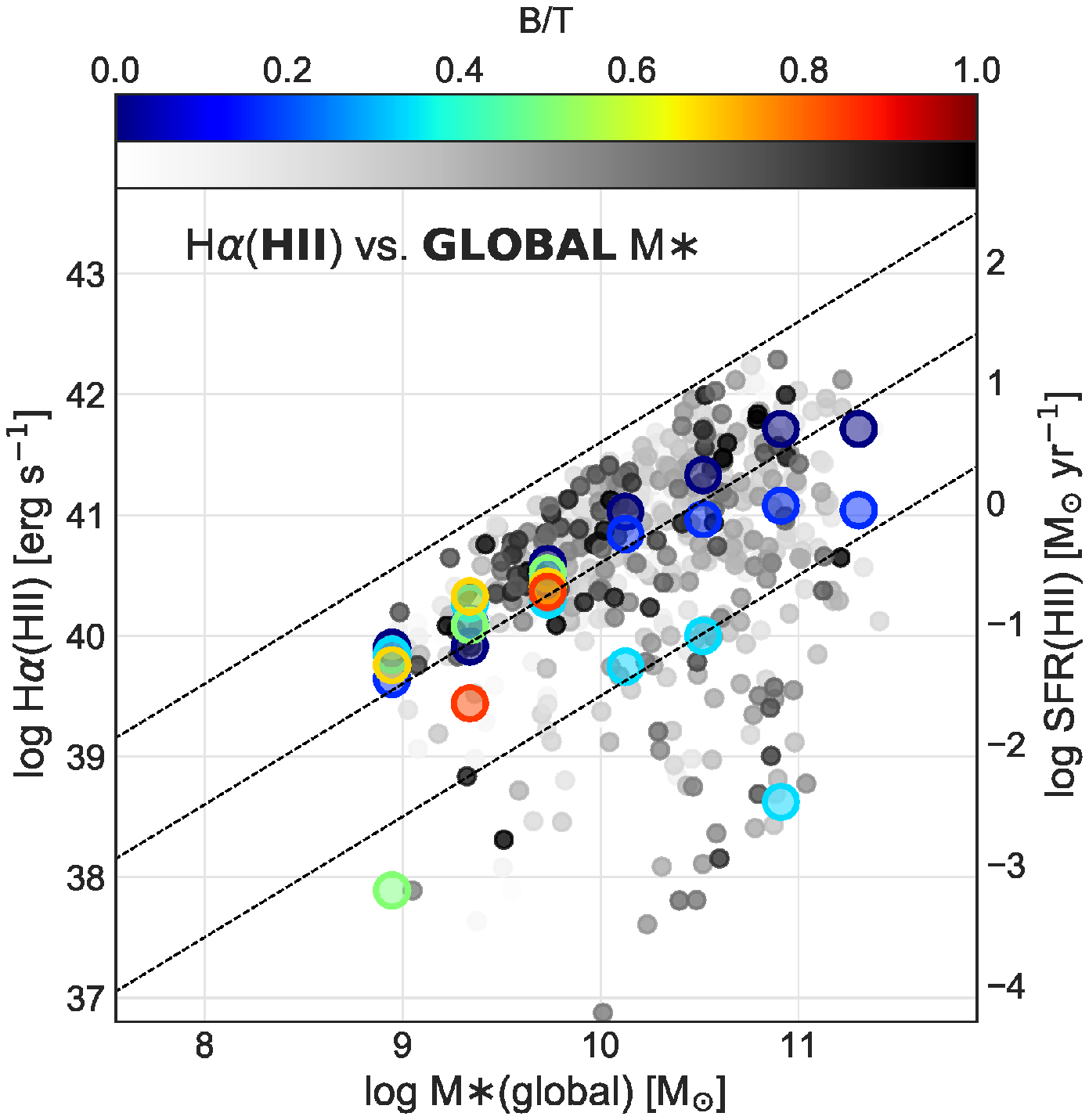}
		\label{fig_global_sm_sfr_hiig2}}
	\subfigure[]{%
	\includegraphics[width=0.33\textwidth]{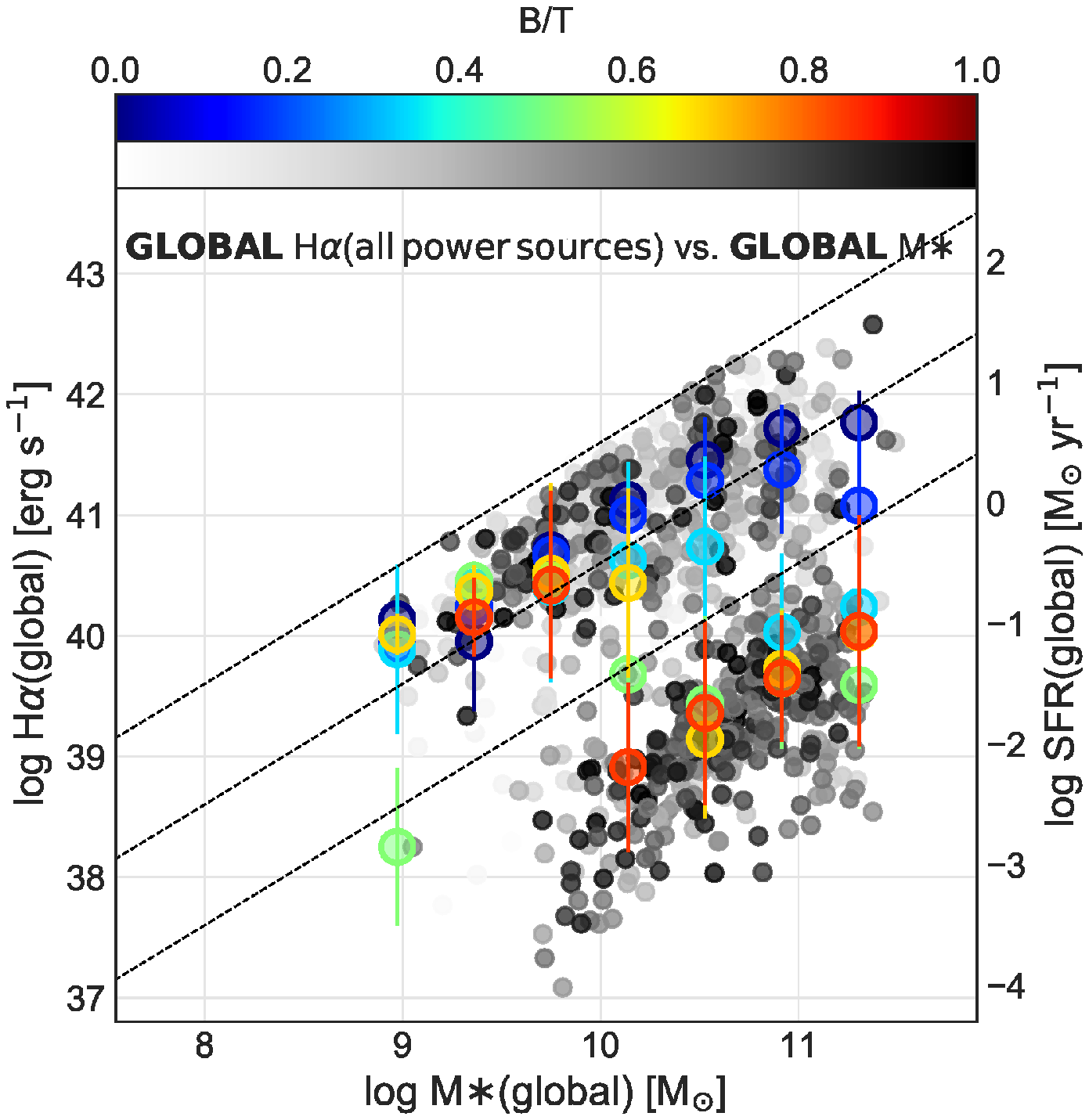}
	\label{fig_global_sm_sfr_all2}}
	\caption{ Integrated  H$\alpha$ and M$_{\ast}$ relations. Individual objects are shown by small circles. Colored circles are the median values of log(H$\alpha$) of the whole sample in different M$_{\ast}$ and B/T bins. The error bars are given by the standard deviation in each bin. The dashed lines represent  log(sSFR/yr$^{-1}$) of -9.5, -10.5, and -11.5  (from top to bottom). (a) Integrated H$\alpha$(HII)-M$_{\ast}$(HII) relation, where  H$\alpha$(HII) and M$_{\ast}$(HII) represent the H$\alpha$ luminosity and stellar mass integrated over only HII spaxels in galaxies.  In other words, the figure represents the integrated relation for star-forming regions.  (b) Integrated  H$\alpha$(HII) versus M$_{\ast}$(global) relation. Since the zero-H$\alpha$(HII)  are taken into account when computing the median H$\alpha$(HII) (color circles), the standard deviation become meaningless due to the highly skewed H$\alpha$(HII) distribution.  The error bars are thus omitted in this plot. (c) Traditional integrated H$\alpha$(global)-M$_{\ast}$(global) relation, same as Figure \ref{fig_global_sm_sfr_integrated}.}
	\label{fig_global_all} 
\end{figure*}

How does the integrated H$\alpha$-M$_{\ast}$ relation look like after excluding the non-HII contribution?
Figure \ref{fig_global_all}  shows the integrated H$\alpha$(HII)-M$_{\ast}$(HII) relation in panel (a),  the  H$\alpha$(HII)-M$_{\ast}$(global) relation in panel (b), and 
the traditional H$\alpha$(global)-M$_{\ast}$(global) relation in panel (c) (same as Figure \ref{fig_global_sm_sfr_integrated}),  where  H$\alpha$(HII) and M$_{\ast}$(HII) represent   H$\alpha$  and M$_{\ast}$ integrated over only HII spaxels  in galaxies. 
Symbol styles and colours are as in the Figure \ref{fig_global_sm_sfr_integrated}.
We note here that,  the H$\alpha$(HII) distribution could become highly skewed if  there are extreme values, such as zero value.
In this case, the standard deviation, which is shown as error bars here, would become meaningless.  
This  affects only Figure \ref{fig_sfr_sm_BPThii}, so the error bars are thus omitted in this plot.

Figure \ref{fig_global_sm_sfr_hiih2} represents the integrated relation for star-forming regions. 
When only looking at star-forming regions, the   H$\alpha$(HII)-M$_{\ast}$(HII) relation shows a relatively tight sequence. 
The figure is simply an integrated version of Figure \ref{fig_sfr_sm_BPThii}. 
 As noted before, the spatially-resolved  relations become similar for the bulge and disk when accounting for  only star-forming regions.
This leads naturally to the  tight and close to linear correlation   in the integrated H$\alpha$(HII)-M$_{\ast}$(HII) relation.

We now shift our focus to the H$\alpha$(HII)-M$_{\ast}$(global) relation in Figure \ref{fig_global_sm_sfr_hiig2}.
For reference, the traditional  H$\alpha$(global)-M$_{\ast}$(global) relation is  displayed in Figure \ref{fig_global_sm_sfr_all2}.
The most noticeable feature in Figure \ref{fig_global_sm_sfr_hiig2}  is the emergence of galaxies with low H$\alpha$(HII)-to-M$_{\ast}$(global) ratio (lower than the H$\alpha$(global)-to-M$_{\ast}$(global) ratio of  quiescent galaxies in the traditional relation).
This  population  is  largely comprised of  the quiescent galaxies, which are dominated by non-HII regions. 
Note that since significant fraction of the highest-B/T and highest-mass galaxies  have little to no H$\alpha$ from star formation, the median H$\alpha$(HII) of these populations  drop to close to zero.

It is worth noting also that the strong sequence of quiescent galaxies observed in the traditional relation  becomes scattered in the H$\alpha$(HII)-M$_{\ast}$(global) relation; no clear scaling relation is found between  H$\alpha$(HII) and M$_{\ast}$(global) for this population.  
Such a lack of bimodality in the SFR distribution at a given stellar mass is very similar to that using the SSP-based SFR by \cite{Gon16} and using MAGPHYS\footnote{Multi-wavelength Analysis of Galaxy Physical Properties \citep{Dac08}: http://www.iap.fr/magphys/}  rather than based on H$\alpha$ by \cite{Eal17}.

\begin{figure*}%
	\centering
	\includegraphics[width=0.99\textwidth]{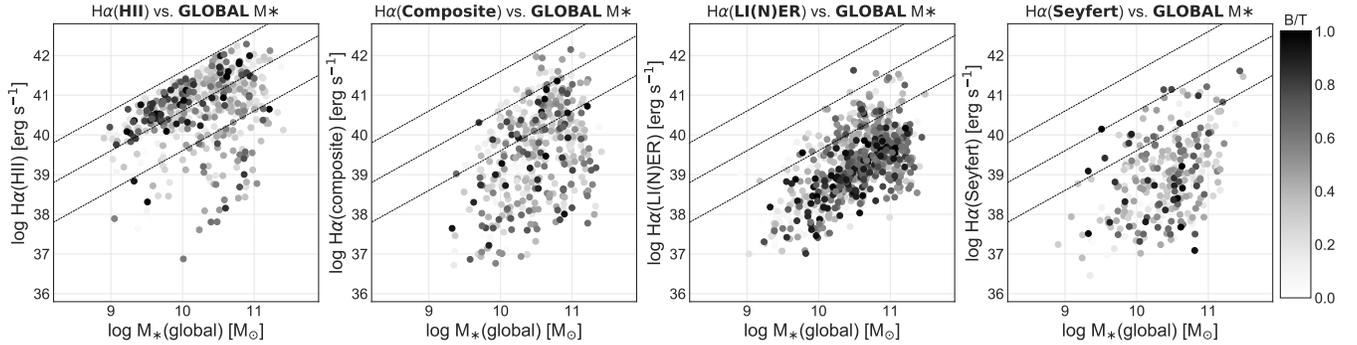}
	\caption{Integrated H$\alpha$(HII), H$\alpha$(composite), H$\alpha$(LI(N)ER), and H$\alpha$(Seyfert)  against M$_{\ast}$(global). From left to right, the $y$-axes are H$\alpha$ luminosity integrated over HII (same as Figure \ref{fig_global_sm_sfr_hiig2}), composite, LI(N)ER, and  Seyfert regions in galaxies, respectively. The circles represent individual galaxies, color-coded  by B/T.  For reader to compare the plots with other figures, the corresponding lines of log(sSFR/yr$^{-1}$) $=$ -9.5, -10.5, and -11.5 (from top to bottom) are shown in all panels. }%
	\label{fig_global_sm_sfr_Sub_SM}%
	\vspace{13pt}
\end{figure*}

Then what  processes fundamentally drive the two strong sequences in the traditional H$\alpha$(global)-M$_{\ast}$(global) relation in Figure \ref{fig_global_sm_sfr_integrated}?
From left to right, the four panels in Figure \ref{fig_global_sm_sfr_Sub_SM} respectively present the  H$\alpha$ luminosity integrated over star formation spaxels (same as Figure \ref{fig_global_sm_sfr_hiig2}), composite spaxels, LI(N)ER spaxels, and Seyfert spaxels against M$_{\ast}$(global).
The high H$\alpha$-to-M$_{\ast}$ ratio regime is heavily populated by star formation, while other mechanisms occupy the low ratio regime. The H$\alpha$(composite)-M$_{\ast}$(global) and  H$\alpha$(Seyfert)-M$_{\ast}$(global) relations show relatively large scatter for a given M$_{\ast}$(global), on the other hand, H$\alpha$(LI(N)ER) and M$_{\ast}$(global) appear to be more tightly correlated to each other. The relation is very similar to the quiescent population  in the traditional H$\alpha$(global)-M$_{\ast}$(global) relation.
In other words, H$\alpha$ powered by LI(N)ER is directly correlated with the underlying stellar mass.
This has been explained as the hot, evolved stars as the dominant mechanism powering the H$\alpha$ emission in quiescent galaxies \citep[e.g.,][]{Bel16,Hsi17}. Our Figure \ref{fig_Frac_all_all} is also in line with this scenario.

Finally, we remind the reader  that the non-HII spaxels are not necessarily devoid of star formation, but simply  being dominated by mechanisms other than star formation. 
That is to say, the H$\alpha$(global)-M$_{\ast}$(global) relation and the H$\alpha$(HII)-M$_{\ast}$(global) relation  represent bracketing scenarios as the \emph{true} SFR of the galaxies would be found between H$\alpha$(HII) and H$\alpha$(global). 
Moreover, the flatting (or turnover) of the integrated SFR-M$_{\ast}$ relation would become \emph{more pronounced} as we move from the traditional integrated SFR(global)-M$_{\ast}$(global) relation to the true SFR versus  M$_{\ast}$(global) relation.

\section{Summary}
\label{sec_summary}

In this work, we present the analysis of the global and spatially-resolved H$\alpha$-M$_{\ast}$ relations using a sample of $\sim$ 1000 galaxies from the MaNGA survey (\S\ref{sec_data}). 
By virtue of the spatially-resolved spectroscopic data from MaNGA,  we spatially identified the regions ionized by different physical processes in each galaxy (\S\ref{sec_results}). 
Our main conclusions are summarized below.

\begin{enumerate}
	\item When all H$\alpha$ powering mechanisms are considered, the spatially-resolved  $\Sigma_{\mathrm{H\alpha}}$(all)-$\Sigma_{\ast}$(all) relation of bulges  progressively turns over  to below the disk sequence  for increasing values of   M$_{\ast}$(global) and/or B/T (Figure \ref{fig_sfr_sm_BPTall}). 
	At the same time, disk sequence is relatively insensitive to galaxy stellar mass and B/T.
	This in turn leads to the frequently reported flattening of the integrated  H$\alpha$(global)-M$_{\ast}$(global) relation in the literature (Figure \ref{fig_global_sm_sfr_integrated}). 
	
	\item On the other hand, we  find little evidence for  the flattening of both integrated H$\alpha$(HII)-M$_{\ast}$(HII) and spatially-resolved $\Sigma_{\mathrm{H\alpha}}$(HII)-$\Sigma_{\ast}$(HII) relations when the star-forming  regions alone are considered (Figure \ref{fig_sfr_sm_BPThii} and \ref{fig_global_sm_sfr_hiih2}).
	
	\item The fractional contribution of  non-HII  sources to  total H$\alpha$ luminosity of a galaxy increases  with increasing B/T and  M$_{\ast}$(global), and increases from disk to bulge regions, suggesting a decreasing role of star formation as an ionizing source toward high-mass, high-B/T galaxies and bulge regions (\S\ref{sec_barplots} and Figure \ref{fig_Frac_all_all}). Moreover, the non-HII  sources tend to have lower ionizing ability compared to star formation.
	
	\item We  discussed the difference between the traditional H$\alpha$(global)-M$_{\ast}$(global) relation and H$\alpha$(HII)-M$_{\ast}$(global) relation (\S\ref{sec_integrated} and Figure \ref{fig_global_all}). There is no clear scaling relation between H$\alpha$(HII) and M$_{\ast}$(global) for the quiescent population. The strong quiescent sequence  in the traditional H$\alpha$(global)-M$_{\ast}$(global) relation is primarily driven by LI(N)ER emissions as shown by  Figure \ref{fig_global_sm_sfr_Sub_SM} and \citet{Hsi17}.

\end{enumerate}

Taken all together, our results imply that   the appearance of galaxy SFR-M$_{\ast}$ relation critically depends on the global properties of galaxies (e.g., stellar mass and B/T)  and relative abundances of various ionizing sources within the galaxies.
The results also emphasize the necessity of spatially-resolved spectroscopy  to understand the origin of galaxy SFR-M$_{\ast}$ relation.

\section*{Acknowledgment}

We thank the anonymous referee for the constructive comments which improved the paper.
The work is supported by the Ministry of Science \& Technology of Taiwan under the grant MOST 103-2112-M-001-031-MY3 and 106-2112-M-001-034-. SFS thanks the CONACyt programs CB-180125 and DGAPA-PAPIIT IA101217 grants for their support to this project.  MB was supported by MINEDUC-UA project, code ANT 1655.
This project makes use of the MaNGA-Pipe3D data products. We thank the IA-UNAM MaNGA team for creating this catalog, and the ConaCyt-180125 project for supporting them.

Funding for the Sloan Digital Sky Survey IV has been provided by
the Alfred P. Sloan Foundation, the U.S. Department of Energy Office of
Science, and the Participating Institutions. SDSS-IV acknowledges
support and resources from the Center for High-Performance Computing at
the University of Utah. The SDSS web site is www.sdss.org.
SDSS-IV is managed by the Astrophysical Research Consortium for the 
Participating Institutions of the SDSS Collaboration including the 
Brazilian Participation Group, the Carnegie Institution for Science, 
Carnegie Mellon University, the Chilean Participation Group, the French Participation Group, Harvard-Smithsonian Center for Astrophysics, 
Instituto de Astrof\'isica de Canarias, The Johns Hopkins University, 
Kavli Institute for the Physics and Mathematics of the Universe (IPMU) / 
University of Tokyo, Lawrence Berkeley National Laboratory, 
Leibniz Institut f\"ur Astrophysik Potsdam (AIP),  
Max-Planck-Institut f\"ur Astronomie (MPIA Heidelberg), 
Max-Planck-Institut f\"ur Astrophysik (MPA Garching), 
Max-Planck-Institut f\"ur Extraterrestrische Physik (MPE), 
National Astronomical Observatories of China, New Mexico State University, 
New York University, University of Notre Dame, 
Observat\'ario Nacional / MCTI, The Ohio State University, 
Pennsylvania State University, Shanghai Astronomical Observatory, 
United Kingdom Participation Group,
Universidad Nacional Aut\'onoma de M\'exico, University of Arizona, 
University of Colorado Boulder, University of Oxford, University of Portsmouth, 
University of Utah, University of Virginia, University of Washington, University of Wisconsin, 
Vanderbilt University, and Yale University.

\end{document}